\documentclass[journal,onecolumn]{IEEEtran}    

\IEEEoverridecommandlockouts                              

\usepackage{amsmath} 
\usepackage{amssymb}  
\usepackage{subcaption}
\usepackage{float}
 
\usepackage[short]{optidef}
\usepackage{pgf}
\usepackage[utf8x]{inputenc}
\usepackage{cite}
\usepackage[colorlinks=false]{hyperref}  

\title{\LARGE \bf
Self-Optimizing Grinding Machines using Gaussian Process Models and Constrained Bayesian Optimization
}

\author{Markus Maier$^{1}$, Alisa Rupenyan$^{1,2}$, Christian Bobst$^{3}$ and Konrad Wegener$^{4}$
\thanks{$^{1}$ The authors are with Inspire AG, Zurich, Switzerland.
        {\tt\small \{maier,rupenyan\}@inspire.ethz.ch}}
\thanks{$^{2}$The author is with the Automatic Control Laboratory, Department of Electrical Engineering and Information Technology, ETH Zurich, Switzerland.
        }%
\thanks{$^{3}$The author is with Agathon AG, Bellach, Switzerland.
        {\tt\small christian.bobst@agathon.ch}}
 \thanks{$^{4}$The author is with Institute for Machine Tools, Department of Mechanical and Process Engineering, ETH Zurich, Switzerland.
        {\tt\small wegener@iwf.mavt.ethz.ch}}
}

\begin{document}

\maketitle

\begin{abstract}
In this study, self-optimization of a grinding machine is demonstrated with respect to production costs, while fulfilling quality and safety constraints. The quality requirements of the final workpiece are defined with respect to grinding burn and surface roughness, and the safety constrains are defined with respect to the temperature at the grinding surface. Grinding temperature is measured at the contact zone between grinding wheel and workpiece using a pyrometer and an optical fiber, which is embedded inside the rotating grinding wheel. Constrained Bayesian optimization combined with Gaussian process models is applied to determine the optimal feed rate and cutting speed of a cup wheel grinding machine manufacturing tungsten carbide cutting inserts. The approach results in the determination of optimal parameters for unknown workpiece and tool combinations after only a few grinding trials. It also incorporates the uncertainty of the constraints in the prediction of optimal parameters by using stochastic process models\footnote{This is a pre-print of an article published in The International Journal of Advanced Manufacturing Technology. The final authenticated version is available online with a DOI 10.1007/s00170-020-05369-9.}.
\end{abstract}
\section{Introduction}
\label{intro}
The selection of optimal grinding parameters is an important task because grinding is often one of the last manufacturing steps where failures lead to high financial losses. Today parameter selection is mainly performed by experienced operators using a trial and error approach. Self-optimization of grinding machines has been the subject of extensive research in industry and academia. The main challenge in grinding process optimization arises from the non-linear, stochastic and transient process properties. Already in 1994, Rowe et al. \cite{Ref1} reviewed artificial intelligence systems applied to grinding and listed various approaches such as knowledge-based systems, expert systems, fuzzy logic systems, neural net systems, and adaptive control optimization techniques. Over time, various models have been proposed to control specific grinding tasks. For example, in \cite{Ref2} a state-space model for the prediction of part size and surface roughness in cylindrical grinding is proposed and \cite{Ref3} reports an adaptive grinding force control. Modelling is strongly coupled with optimization. An overview on grinding process modelling and optimization is given in \cite{Ref4}. For example, in \cite{Ref5} a method for the optimization of a grinding process in batch production using a grinding process model and a dynamic programming approach to update process parameters between grinding cycles is demonstrated. In \cite{Ref6} constrained multi-objective optimization of grinding with Pareto fronts using genetic algorithms is shown. Another ingredient for optimization is process monitoring using measurements from force or power sensors, acoustic emission sensors and temperature sensors \cite{Ref7}. The combination of models, optimization techniques, sensor data, and material data enables intelligent, data-driven methods for process control and assistance, as demonstrated in \cite{Ref8,Ref9,Ref10}. In \cite{Ref8} the grinding cycles are adaptively optimized and grinding burn is prevented, \cite{Ref9} presents empirical models fitted with experimental data, used to find optimal process parameters based on optimization objective minimization subjected to constraints, and \cite{Ref10} describes a model-based assistant system for precision, productivity, and stability increase of centerless grinding.

Current state of the art methods work well for specific tasks and niche applications where extensive training data is available or low model complexity is sufficient. However, industrial applications for general process optimization in grinding are still missing. Grinding processes are very sensitive to environmental conditions, and optimizing a grinding process on one machine does not necessarily result in the same performance on a different machine of the same type. Another obstacle is the lack of information such as physical properties of the grinding wheel and the workpiece because they are trade secrets of the respective manufacturer. Therefore, on-machine experiments are needed for optimization such as done today by experienced operators. Experimental design methods, such as the Taguchi method, can be used to determine on-machine experiments, as presented in \cite{Ref11}. The disadvantage of the Taguchi method is that knowledge gained during testing is not incorporated in the optimization.

In the present study, Gaussian process (GP) regression is used for process modelling. Gaussian process regression is a powerful stochastic model, which is used widely to provide uncertainty estimation when dealing with stochastic data \cite{Ref12}. The main advantage of a Gaussian process regression is its flexibility, the in-built uncertainty prediction, and the non-parametric nature of the model. A detailed introduction to Gaussian process regression can be found in \cite{Ref12}. Based on a Gaussian process prior and available measurements, Gaussian process regression can be used to predict the mean and the variance for every arbitrary combination of input parameters. The estimated variance is useful to incorporate uncertainty in the process optimization based on the predictions from the Gaussian process regression.

Gaussian process regression is used in combination with Bayesian optimization (BO), where it provides a surrogate model for the unknown objective function. Bayesian optimization is a sequential design strategy for global optimization of an unknown objective function, which is expensive to evaluate \cite{Ref13}. Bayesian optimization uses the predicted mean and variance functions of the Gaussian process model of the objective to determine the next experimental point with high information content, afterwards the GP model is updated with the newly available data \cite{Ref13}. This procedure is repeated until suitable process parameters are found. Bayesian optimization has been successfully applied in hyperparameter tuning of learning algorithms \cite{Ref14}, photovoltaic power plants \cite{Ref15}, and robotics \cite{Ref16}. Recently, it has also been successfully applied to the optimization of turning processes \cite{Ref17,Ref18}.

In this study, an autonomous process set-up is proposed where the cutting speed and the feed rate in grinding are optimized based on combined measurements of roughness, temperature, and maximum dressing intervals. Constrained Bayesian optimization is applied to enforce quality and safety constraints. This paper is organized as follows: A short introduction to Bayesian optimization and Gaussian process models is provided first, followed by a methodology section explaining the optimization objective, experimental set-up, and algorithm implementation. Afterwards, optimization results performed on a cup wheel grinding process are reported and discussed.

\section{Bayesian optimization and Gaussian process models}
\label{sec:2}
A general introduction to Gaussian process models is given in \cite{Ref12} and an introduction to Bayesian optimization can be found in \cite{Ref13}. According to \cite{Ref12}, a Gaussian process represents a distribution over functions and is a collection of random variables, which have a joint Gaussian distribution. A Gaussian process is fully defined by a mean function $m(x)$ and a covariance function $k(x,x')$ often referred as the kernel. In this study, a Matern 5 kernel with automatic relevance determination (ARD) \cite{Ref12} is used,
\begin{equation}
\label{eqn:1}
k(\mathbf{x},\mathbf{x}')=\sigma_f^2 \left(1+\sqrt{5} r+\frac{5}{3} r^2 \right)  \exp-\sqrt{5} r) 
\end{equation}

\begin{equation}
\label{eqn:2}
r=\sqrt{(\mathbf{x}-\mathbf{x}')^T\mathbf{\mathbf{P}}^{-1}(\mathbf{x}-\mathbf{x}')}
\end{equation}

\begin{equation}
\label{eqn:3}
\mathbf{\mathbf{P}}=\mathrm{diag}(l_1^2,l_2^2,...,l_D^2)
\end{equation}
where $\sigma_f^2$ is the signal variance, $r$ is the distance between the input data points $\mathbf{x}$ and $\mathbf{x}'$ (in this case, the input data are process parameters that have to be optimized), and $\mathbf{\mathbf{P}}$ is a diagonal matrix containing characteristic length scale parameters $l_i^2$ for each input space dimension up to dimension $D$. The kernel function specifies the relation between the function values at process parameters $\mathbf{x}$ and $\mathbf{x}'$, where for a short distance between the process parameters the corresponding function values are similar whereas for larger distances higher variations are observed.

A Gaussian process is used as a prior for Gaussian process regression. In this study, the Gaussian process prior mean function $m(\mathbf{x})$ is assumed zero, specifying no expert knowledge. The resulting posterior mean $\mu_t$ and variance $\sigma_t^2$ of  the Gaussian process regression  at an arbitrary location $\mathbf{x}$ can be calculated as follows \cite{Ref12},

\begin{equation}
\label{eqn:4}
\mu_t(\mathbf{x})=\mathbf{k}^T(\mathbf{x})(\mathbf{\mathbf{K}}+\sigma^2_N\mathbf{\mathbb{I}})^{-1}\mathbf{y}_t
\end{equation}

\begin{equation}
\label{eqn:5}
\sigma_t^2(\mathbf{x})=k(\mathbf{x},\mathbf{x})-\mathbf{k}^T(\mathbf{x})(\mathbf{\mathbf{K}}+\sigma^2_N\mathbf{\mathbb{I}})^{-1}\mathbf{k}(\mathbf{x})
\end{equation}

\begin{equation}
\label{eqn:6}
\mathbf{\mathbf{K}} = 
\left( \begin{array}{ccc} 

k(\mathbf{x}_1,\mathbf{x}_1) & \cdots & k(\mathbf{x}_1,\mathbf{x}_t) \\
\vdots & \ddots & \vdots  \\
k(\mathbf{x}_t,\mathbf{x}_1) & \cdots & k(\mathbf{x}_t,\mathbf{x}_t) 

\end{array}\right)
\end{equation}
where $\mathbf{x}_t$ and $\mathbf{y}_t$ are the vectors containing $t$ pairs of measurements assumed to be corrupted with Gaussian noise $N(0,\sigma_N^2)$, $\mathbf{\mathbb{I}}$ is the identity matrix, $\mathbf{\mathbf{K}}$ is the covariance matrix, and $\mathbf{k}(\mathbf{x})$ is a covariance vector between $\mathbf{x}$ and $\mathbf{x}_1$ to $\mathbf{x}_t$.

\begin{figure}
\center
  \includegraphics[width=0.6\textwidth]{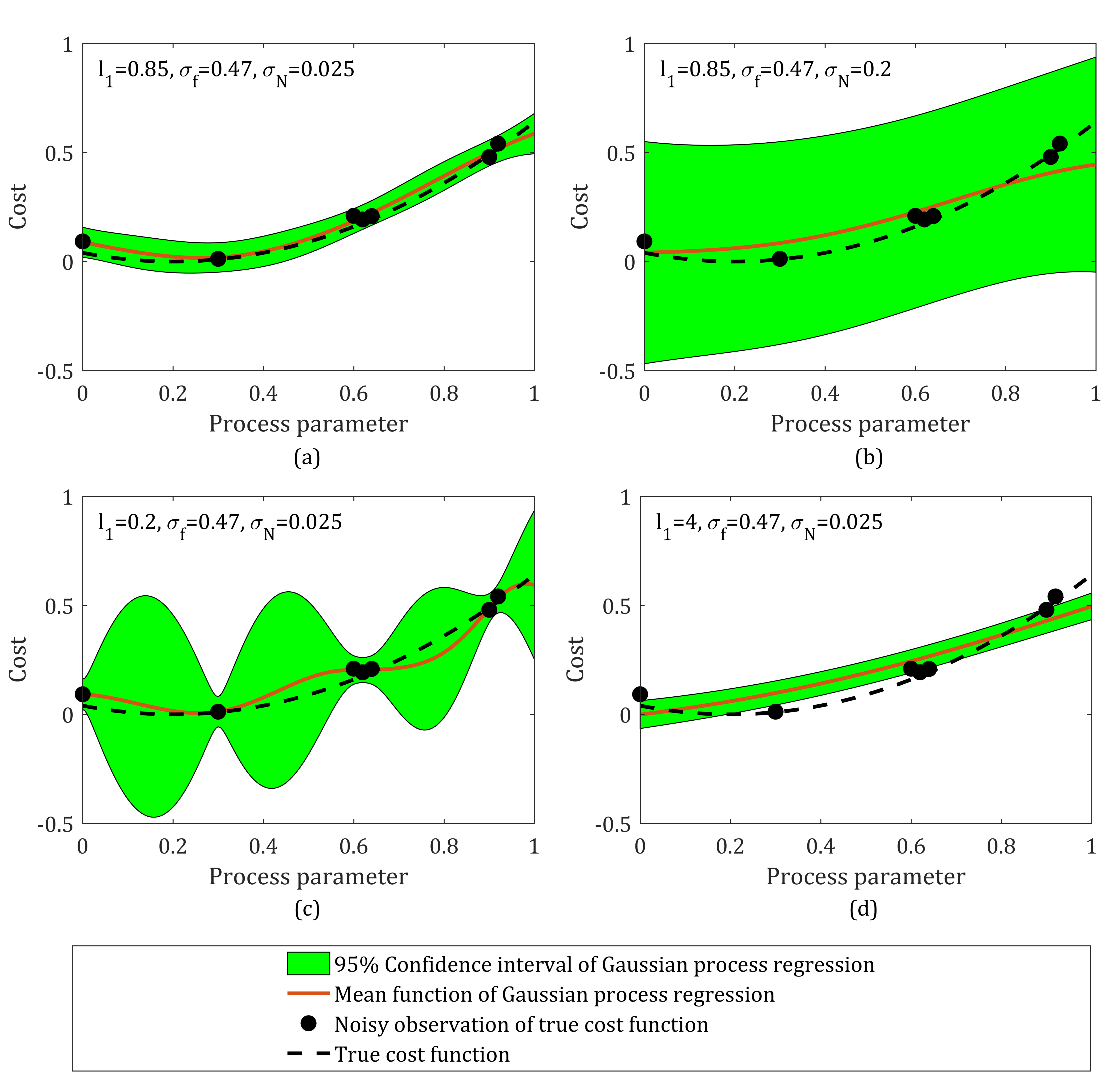}
\caption{Gaussian process regression for different hyperparameters. Panel (a): hyperparameters determined by maximizing the marginal log likelihood using equation (\ref{eqn:8}). Panel (b): results for high signal noise. Panel (c): results for short length scale parameter. Panel (d): results for long length scale parameter.}
\label{fig:1}       
\end{figure}

The length scales $l_i^2$ in eq.  (\ref{eqn:3}), the signal variance $\sigma_f^2$ in eq. (\ref{eqn:1}), and the signal noise $\sigma_N^2$ in eq. (\ref{eqn:4}) and eq. (\ref{eqn:5}) are hyperparameters of the Gaussian processes regression, which can be summarized in a hyperparameter vector $\mathbf{\theta}$. It is possible to specify these hyperparameters before the Gaussian process regression based on expert knowledge. Another method, as described in \cite{Ref12}, is to determine the hyperparameters $\mathbf{\theta}^*$ based on available measurements by maximization of the marginal log likelihood $p(\mathbf{y}_t |\mathbf{\theta})$.

\begin{equation}
\label{eqn:7}
\log p(\mathbf{y}_t|\mathbf{\theta})=-\frac{1}{2}\mathbf{y}^T_t(\mathbf{\mathbf{K}}_{\mathbf{\theta}}+\sigma^2_N\mathbf{\mathbb{I}})^{-1}\mathbf{y}_t-\frac{1}{2}\log |\mathbf{\mathbf{K}}_{\mathbf{\theta}}+\sigma^2_N\mathbf{\mathbb{I}}|-\frac{t}{2}\log 2\pi
\end{equation}

\begin{equation}
\label{eqn:8}
\mathbf{\theta}^*=\mathrm{argmax} \log p(\mathbf{y}_t|\mathbf{\theta})
\end{equation}
A $\mathbf{\theta}$ subscript is added to the covariance matrix in equation (\ref{eqn:7}) to explicitly show its dependence on the hyperparameters $\mathbf{\theta}$. Figure \ref{fig:1} shows the Gaussian process regression results exemplarily for different hyperparameters. In the example the true cost function is set to $y=(x-0.2)^2$ and displayed for illustration (dotted black line). The true cost function was evaluated at several process parameter values. The measurements are noisy observations of the true function $\bar{y}=y+\epsilon$, where $\epsilon \sim N(0,5e-4)$. These measurements are the input data points to the Gaussian process regression. Figure \ref{fig:1}(a) shows the Gaussian process regression for the hyperparameters maximizing the marginal log likelihood using equation (\ref{eqn:8}). It can be seen that with these hyperparameters the fit of the data is very good. A GP regression using the same data points but a high signal noise hyperparameter is shown in Figure \ref{fig:1}(b). The prediction of the true cost function is still reasonable but shows a high overall uncertainty. Figure \ref{fig:1}(c) shows the Gaussian process regression for a low length scale hyperparameter. In this case more complex functions are also plausible candidates, which leads to high uncertainties between the measurements, where data is missing. On the other hand on Figure \ref{fig:1}(d) a Gaussian process regression for a very long length scale hyperparameter is shown. In this case the generalization from the observed measurements results in a simple (nearly linear) function for the mean function of the Gaussian process and in low uncertainty over the whole range. In this case the true cost function is not always within the 95\% confidence interval of the Gaussian process regression due to the predominant generalization. As pointed out in \cite{Ref12}, an advantage of maximizing the marginal log likelihood to determine the hyperparameters is that it naturally trades off model complexity and goodness of model fit.

\begin{figure}
\center
  \includegraphics[width=0.6\textwidth]{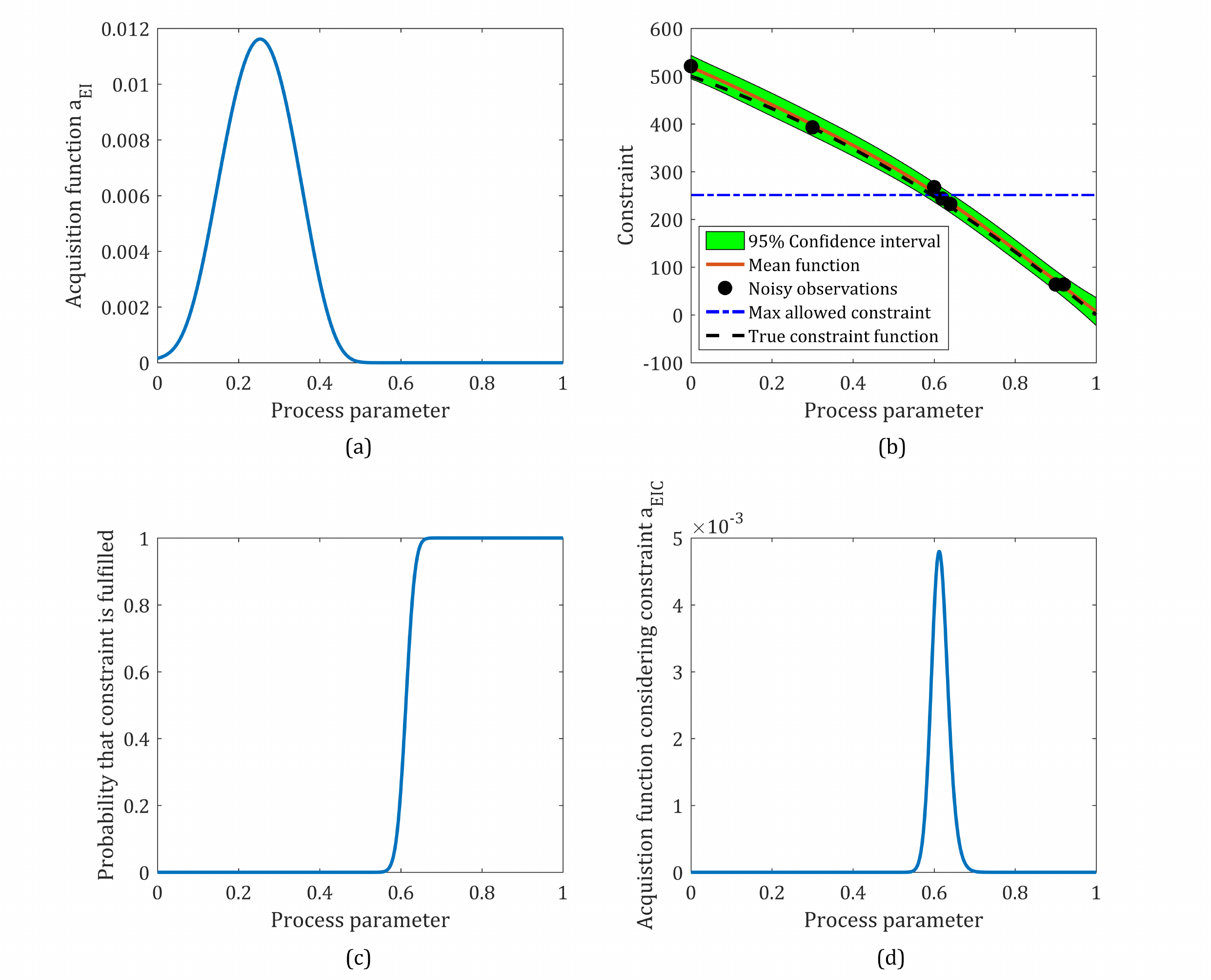}
\caption{Panel (a) shows the calculation of the acquisition function without considering constraints using equation (\ref{eqn:9}) and the Gaussian process regression model of the cost as displayed in Figure \ref{fig:1}(a). Panel (b) shows the Gaussian process regression model of the constraint exemplarily, calculated by maximizing the marginal log likelihood. Panel (c) shows the probability that the constraint is fulfilled, calculated using equation (\ref{eqn:12}) and the Gaussian process regression of the constraint, as displayed in panel (b). Panel (d) shows the result of the acquisition function considering constraints, which can be calculated using equation (\ref{eqn:11}).}
\label{fig:2}       
\end{figure}

Once a GP model of the objective is available, in Bayesian optimization an acquisition function is maximized to determine the candidate parameters where the next experimental trial is conducted. In general the acquisition function provides a trade off between process parameters which are associated with a high uncertainty (exploration), and process parameters with a low predicted value (exploitation) \cite{Ref13}. A common used acquisition function is the expected improvement function, which for a minimization problem can be calculated as follows \cite{Ref19},

\begin{equation}
\label{eqn:9}
a_{EI}(\mathbf{x})=(C_{min}-\mu_{t,co}(\mathbf{x}))F(Z)+\sigma_{t,co}(\mathbf{x})\phi(Z)
\end{equation}

\begin{equation}
\label{eqn:10}
Z=\frac{C_{min}-\mu_{t,co}(\mathbf{x})}{\sigma_{t,co}(\mathbf{x})}
\end{equation}
where $\mu_{t,co}(\mathbf{x})$ is the predicted mean cost function after $t$ iterations obtained by Gaussian process regression using equation (\ref{eqn:4}), $\sigma_{t,co}^2 (\mathbf{x})$ is the predicted variance of the cost after $t$ iterations obtained by Gaussian process regression using equation (\ref{eqn:5}), $C_{min}$ is the so far lowest measured cost, $F(Z)$ is the cumulative standard normal distribution $F(Z)=1/\sqrt{2\pi}\int_{-\infty}
^{Z}\exp-t^2/2)dt$, and $\phi(Z)$ is the probability density function of a standard normal distribution $\phi(Z)=1/\sqrt{2\pi}\exp(-Z^2/2)$. A standard normal distribution is a normal distribution with zero mean and standard deviation of one. The acquisition function is displayed exemplarily for a cost minimization problem in Figure~\ref{fig:2}(a) using the Gaussian process regression from Figure \ref{fig:1}(a). Based on the acquisition function the next experiment is conducted at a process parameter of 0.25, which maximizes the acquisition function.

Most machining operations require fulfilling constraints such as workpiece quality, process or safety constraints. As proposed in \cite{Ref20}, the expected improvement acquisition function can be extended to constrained optimization, which is used in this study,
\begin{equation}
\label{eqn:11}
a_{EIC}(\mathbf{x})=a_{EI}(\mathbf{x})\prod_{i=1}^{n} p_{f,i}(\mathbf{x})
\end{equation}
where $a_{EI} (\mathbf{x})$ is the expected improvement without constraints (\ref{eqn:9}), $n$ is the number of constraints, and $p_{f,i}$ is the probability that the constraint $i$ is fulfilled. Note that for the constrained case the best observed value $C_{min}$ is the minimal measured cost value, which fulfills all constraints - and not the absolute minimal measured cost as in the case without constraints.  The probability that the constraint $i$ is fulfilled can be calculated as follows,
\begin{equation}
\label{eqn:12}
p_{f,i}(\mathbf{x})=\frac{1}{\sigma_{t,c_i}\sqrt{2\pi}}\int_{-\infty}^{\lambda_i}\exp\left(\frac{-(t-\mu_{t,c_i}(\mathbf{x}))^2}{2\sigma^2_{t,c_i}(\mathbf{x})}\right) dt
\end{equation}
where $\lambda_i$ is the maximum allowed value for constraint $i$, $\mu_{t,c_i}(\mathbf{x})$ is the predicted mean of constraint $i$ after $t$ iterations calculated using (\ref{eqn:4}), and $\sigma_{t,c_i}^2(\mathbf{x})$ is the variance of the constraint $i$ after $t$ iterations calculated using (\ref{eqn:5}). Here, the constraint $i$ is modeled again using Gaussian process regression and the predicted mean and uncertainty are used in the calculation. The next experiment is conducted for process parameters $\mathbf{x}^*$, which maximize the constrained expected improvement acquisition function.
\begin{equation}
\label{eqn:13}
\mathbf{x}^*=\mathrm{argmax}\, a_{EIC}(\mathbf{x})
\end{equation}

Figure \ref{fig:2}(b) shows exemplarily the result of a Gaussian process regression for a constraint quantity. In a real application, the quantity might be surface roughness of the workpiece or the temperature during the grinding process and must stay below a specified limit, which depends on the requirements such as the final workpiece quality. In this example,the limit is set to 250. The probability that the constraint is fulfilled is displayed in Figure \ref{fig:2}(c) and can be calculated using equation (\ref{eqn:12}) and the Gaussian process regression of the constraint, as displayed in Figure~\ref{fig:2}(b). Finally, the constrained expected improvement acquisition function is displayed in Figure \ref{fig:2}(d) and can be calculated according to equation (\ref{eqn:11}), using the Gaussian process regression of the cost (displayed in Figure \ref{fig:1}(a)) and the Gaussian process regression of the constraint (displayed in Figure \ref{fig:2}(b)). The constrained expected improvement acquisition function is maximal for a process parameter value of 0.61, which is the next experimental parameter. The constrained expected improvement acquisition function favors higher process parameters than the acquisition function without considering constraints because according to the Gaussian process regression of the constraint the probability that low process parameters fulfill the maximum allowed constraint of 250 is low. After the next experiment is conducted at a process parameter value of 0.61, the optimization procedure is repeated with the newly available measurements until a maximum number of iterations or a stopping criterion is reached.

\section{Methodology}
\label{sec:3}
\subsection{Grinding operation}
\label{sec:31}
\begin{figure}
\center
  \includegraphics[width=0.4\textwidth]{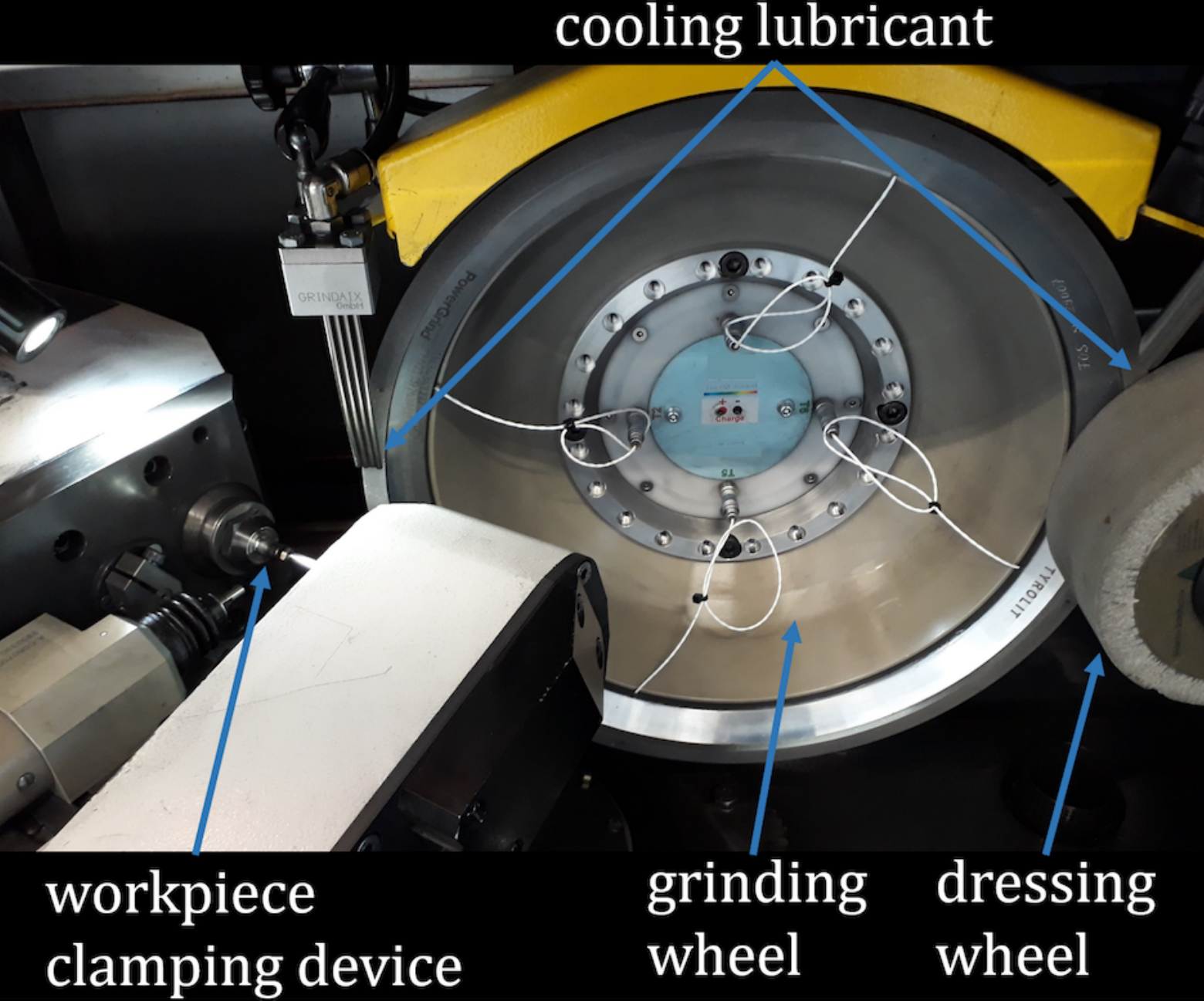}
\caption{Overview of grinding kinematic and experimental setup}
\label{fig:3}       
\end{figure}

The grinding operation is performed on an Agathon DOM Semi grinding machine (see Figure \ref{fig:3}). An initial quadratic shaped insert (14.5 mm x 14.5 mm x 4.76~mm) made of tungsten carbide (Tribo S25, grain size 2.5 $\mu$m, HV30 1470) is ground on two opposite sides, resulting in a final insert geometry of 10 mm x 14.5 mm x 4.76~mm. The machine is equipped with a metal bonded diamond grinding wheel with a mean diamond grain size of 46 $\mu$m and a grain concentration of C100 (D46C100M717 from Tyrolit). During grinding, the grinding wheel is oscillated with 1 Hz to ensure a uniform wear of the grinding wheel. The spark-out time is set to 3 seconds and the oscillation during spark-out is slightly increased to 1.5 Hz. After the grinding wheel is worn out, it is conditioned by a dressing wheel (89A 240 L6 AV217, geometry specification 150x80x32 W12 E15 from Tyrolit). For all experiments the dressing parameters were constant (dressing infeed is 0.1 mm, grinding wheel speed is 6 m/s, dressing wheel speed is 12 m/s, feed rate is 0.5 mm/min, and spark-out time is 1 sec). The used cooling lubricant is Blasogrind HC 5 (from Blaser Swisslube) with a constant flow rate supplied by a multi-needle nozzle as shown in Figure~\ref{fig:3}.

\subsection{Measurement set-up and process constraints}
\label{sec:32}
\begin{figure}
\center
  \includegraphics[width=0.5\textwidth]{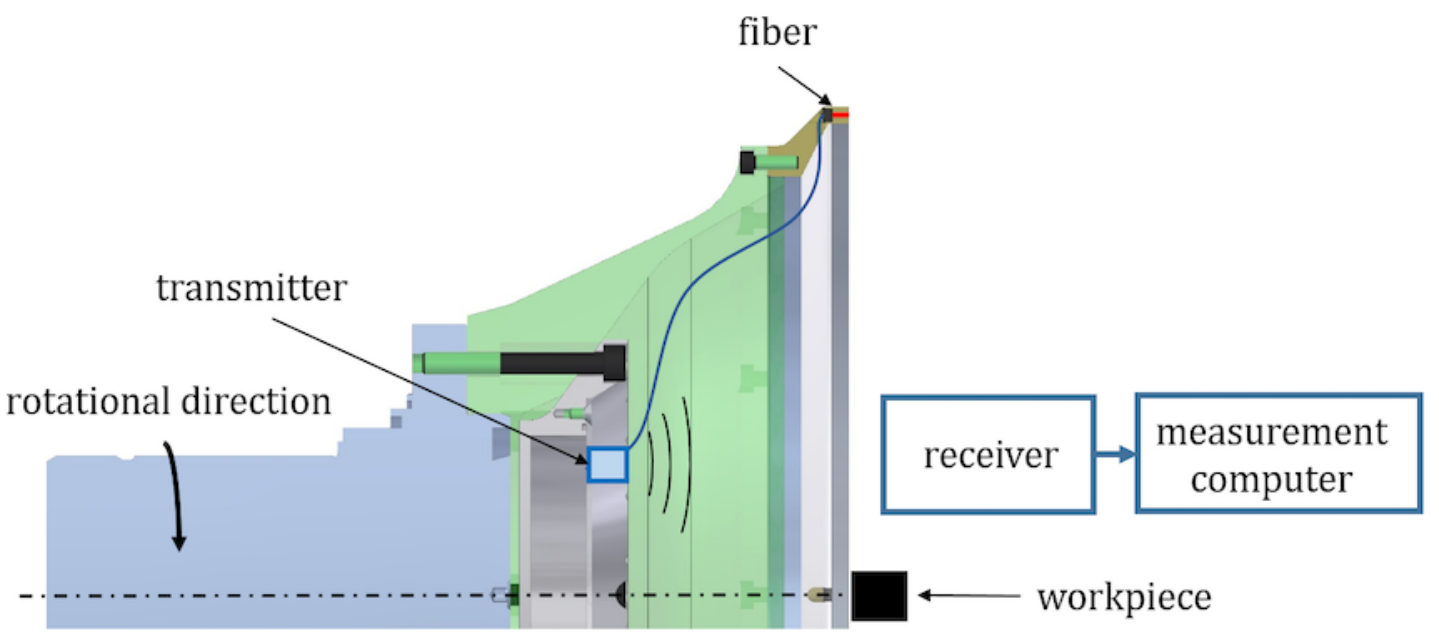}
\caption{Schematic of embedded temperature measurement}
\label{fig:4}       
\end{figure}

\begin{figure}
\center
  \includegraphics[width=0.6\textwidth]{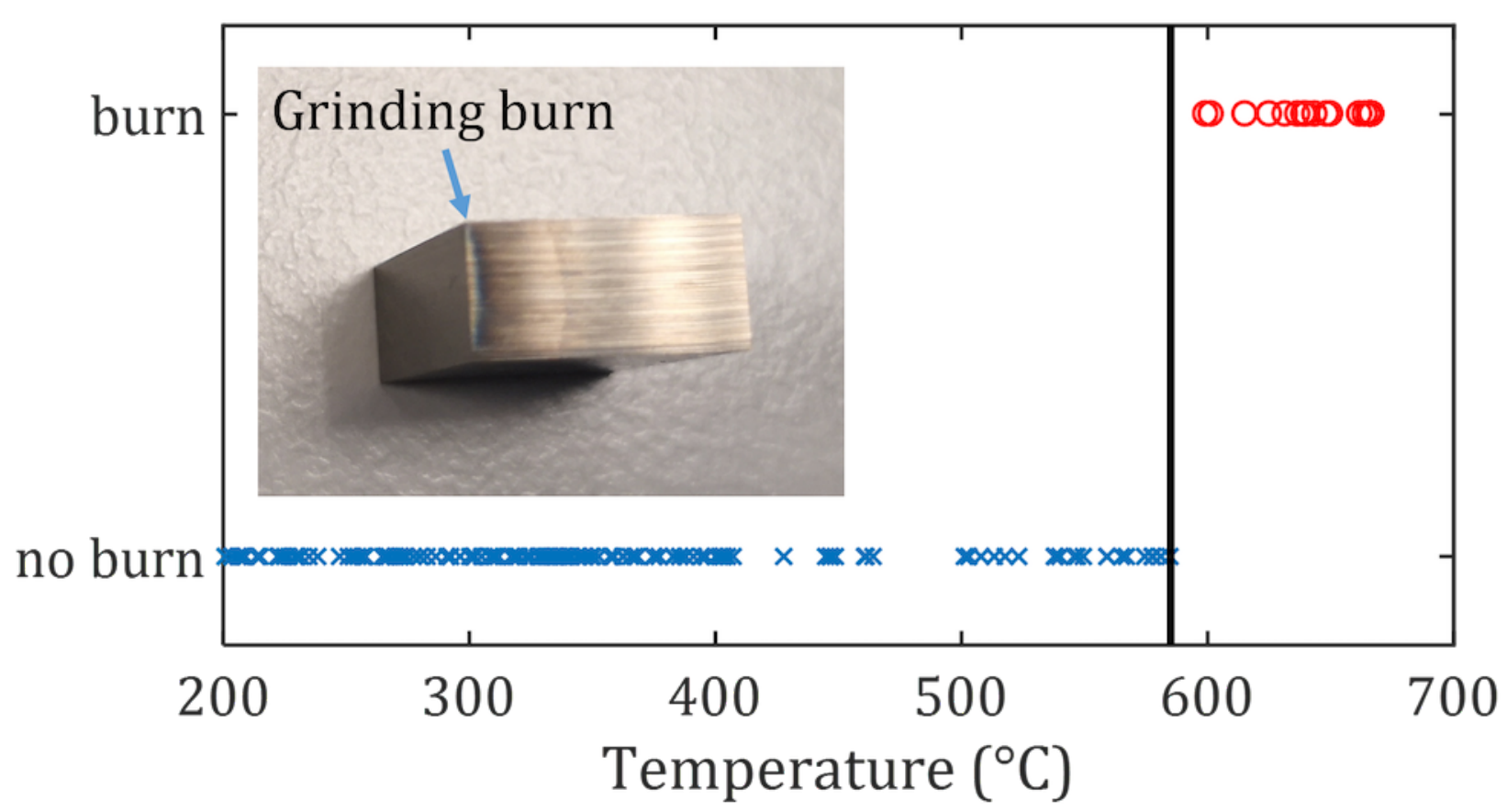}
\caption{Grinding burn limit of preliminary experiments adapted from \cite{Ref21}}
\label{fig:5}       
\end{figure}

High temperatures at the contact zone between grinding wheel and workpiece result in severe thermal damage of the workpiece and in grinding burn. Grinding burn causes an undesired structure change of the workpiece material and must be avoided. In addition, grinding with high temperatures may result in ignition of the grinding oil and results in potential hazardous situations. In this study, grinding burn limited the maximum allowed temperature because grinding burn was observed at lower temperatures than ignition of grinding oil. The temperature is measured optically using an optical fiber embedded inside the grinding wheel, which collects the emitted radiation at the contact zone of the grinding wheel and the workpiece. Figure \ref{fig:4} shows a schematic of the measurement equipment. The optical signal is evaluated on the rotating grinding wheel (pyrometer), filtered and transmitted to a receiver unit outside the grinding machine and converted to temperature. In total, four optical fibers are evenly distributed on the grinding wheel circumference for the temperature measurement. For each grinding operation, the final considered temperature is averaged twice. First, the average of the ten highest temperature readings is calculated, followed by averaging the computed values for each sensor. Note that the evaluated temperature of the pyrometer depends on emissivity of the measured surface. The temperature sensor was not calibrated, which allowed only for measurements of relative temperature. As the temperature sensor is used to detect grinding burn with a burn threshold temperature determined in advance, having access to relative temperatures proved sufficient for the goals of the grinding process optimization. If the first ground insert side after dressing already is exposed to high temperature and grinding burn, the workpiece quality is not fulfilled for the current and for the following workpieces, as in general the grinding temperature increases during grinding due to grinding wheel dulling. The measured temperature of the first ground side after dressing is therefore a constraint in the subsequent optimization procedure and should not exceed the allowed threshold value.
The temperature readings after dressing were used to determine the maximum dressing interval. The maximum dressing interval is defined as the number of grinding cycles until the temperature exceeds the maximum allowed temperature. The experimental run was stopped when this temperature limit was exceeded. To avoid unnecessary long experimental runs, the run was stopped after 8 inserts if the maximum temperature was not previously exceeded. The maximum allowed temperature was set to 585$^{\circ}$C because in preliminary experiments in \cite{Ref21} with 194 insert sides grinding burn has been classified with a 100\% success rate using this temperature measurement system and the same workpiece-wheel combination as in this study. Figure \ref{fig:5} shows the classification of grinding burn based on a temperature threshold and temperature measurements. In \cite{Ref21} grinding burn is detected optically by the black coloring of the workpiece surface. This is a simple method for grinding burn detection which is often used in an industrial environment. In this study, this approach is considered sufficient because it is only used to specify the maximum allowed temperature and thereby the desired workpiece quality. For workpieces with higher quality requirements, the maximum temperature might be determined based on more sophisticated methods, such as Barkhausen noise measurements as applied in \cite{Ref22}. 

The surface roughness of the workpiece is measured transversal to the grinding direction and orthogonal to the cutting edge using a tactile measurement device (Form Talysurf 120 L from Taylor Hobson) with a measurement tip radius of 2 $\mu$m. The evaluation length was reduced to 4 mm because the total thickness of the insert was only 4.76 mm. The wavelength cut-off is set to $\lambda_c$ = 0.8 mm. In general, the roughness of both ground sides per insert is measured and averaged. If grinding burn occurred on a ground side the roughness was not measured on this side. Instead, the roughness measurement was repeated at a different location on the other ground side, where grinding burn did not occur. The roughness measurements change from insert to insert with no clear trend. Therefore, in a conservative approach, the maximum measured roughness value of one optimization run, grinding with a freshly dressed wheel until grinding burn occurred, was considered as a constraint for the optimization. In this study, the roughness was limited to $R_{a,max}$ = 230~nm.

\subsection{Cost calculation}
\label{sec:33}
The grinding operation cost cannot be measured directly with a single sensor as it is composed of both process-related quantities such as grinding time and grinding wear, and of economic quantities such as machine hourly cost and cost of the grinding wheel. The total cost to grind one insert is calculated based on  grinding time cost, dressing time cost, cost of grinding wheel wear during dressing, and cost of dressing wheel wear during dressing.

\begin{equation}
\label{eqn:14}
C_T(v_s,f)=C_m\frac{s\cdot a_{pg}}{f}+\left(C_m t_d+\frac{C_{gw}a_{d,gw}}{a_{gw}}+\frac{C_{dw} a_{d,dw}}{a_{dw}}\right)\frac{V_g}{V_w(v_s,f)}
\end{equation}
The feed rate $f$, and the removed workpiece material until the wheel is dull $V_w$ depend on the selected input parameters feed rate $f$ and cutting speed $v_s$, which are the process parameters used in Bayesian optimization. The other cost parameters are constants as summarized in Table \ref{tab:1}. The grinding wheel wear during dressing was considered and the wear during grinding was neglected. This assumption is reasonable because it is known that metal bonded grinding wheels hold the diamond grains very strongly, resulting in a high abrasive wear resistance whereby self-sharpening is limited \cite{Ref23}. The limited self-sharpening of the grinding wheel requires redressing of the grinding wheel, which is assumed to be the main cause of grinding wheel wear. The grinding wheel wear during dressing $a_{d,gw}$ was assumed to be 1 $\mu$m. Grinding wheel wear during dressing is determined by the dressing parameters, which were fixed in this study and not considered in the optimization. In this study, grinding wheel wear during dressing can be considered as a weight in the cost function.  The spark-out time for each grinding operation was not considered for cost calculation because it only shifts the cost function up by a constant value and does not influence the optimal parameters. For generality, the cost is specified with units U.

\begin{table}
\centering
\caption{Parameters for cost calculation}
\label{tab:1}       
\begin{tabular}{lll}
\hline\noalign{\smallskip}
Parameter & Description & Value  \\
\noalign{\smallskip}\hline\noalign{\smallskip}
$V_g$ & Removed material per workpiece & 310.6 mm\textsuperscript{3} \\
$s$ & Number of ground sides & 2 \\
$a_{pg}$ & Infeed per side & 2.25 mm \\
$t_d$ & Dressing time & 19 sec \\
$a_{gw}$ & Abrasive layer thickness of the grinding wheel & 4 mm \\
$a_{d,gw}$ & Wear of the grinding wheel during dressing & 1 $\mathrm{\mu}$m \\
$a_{dw}$ & Thickness of the dressing wheel & 65 mm \\
$a_{d,dw}$ & Wear of the dressing wheel during dressing & 0.1 mm  \\
$C_m$ & Machine hourly cost & 100 U/h  \\
$C_{gw}$ & Cost of grinding wheel & 1500 U  \\
$C_{dw}$ & Cost of dressing wheel & 95 U  \\
\noalign{\smallskip}\hline
\end{tabular}
\end{table}

\subsection{Optimization Implementation}
\label{sec:34}
\begin{figure}
\center
  \includegraphics[width=0.5\textwidth]{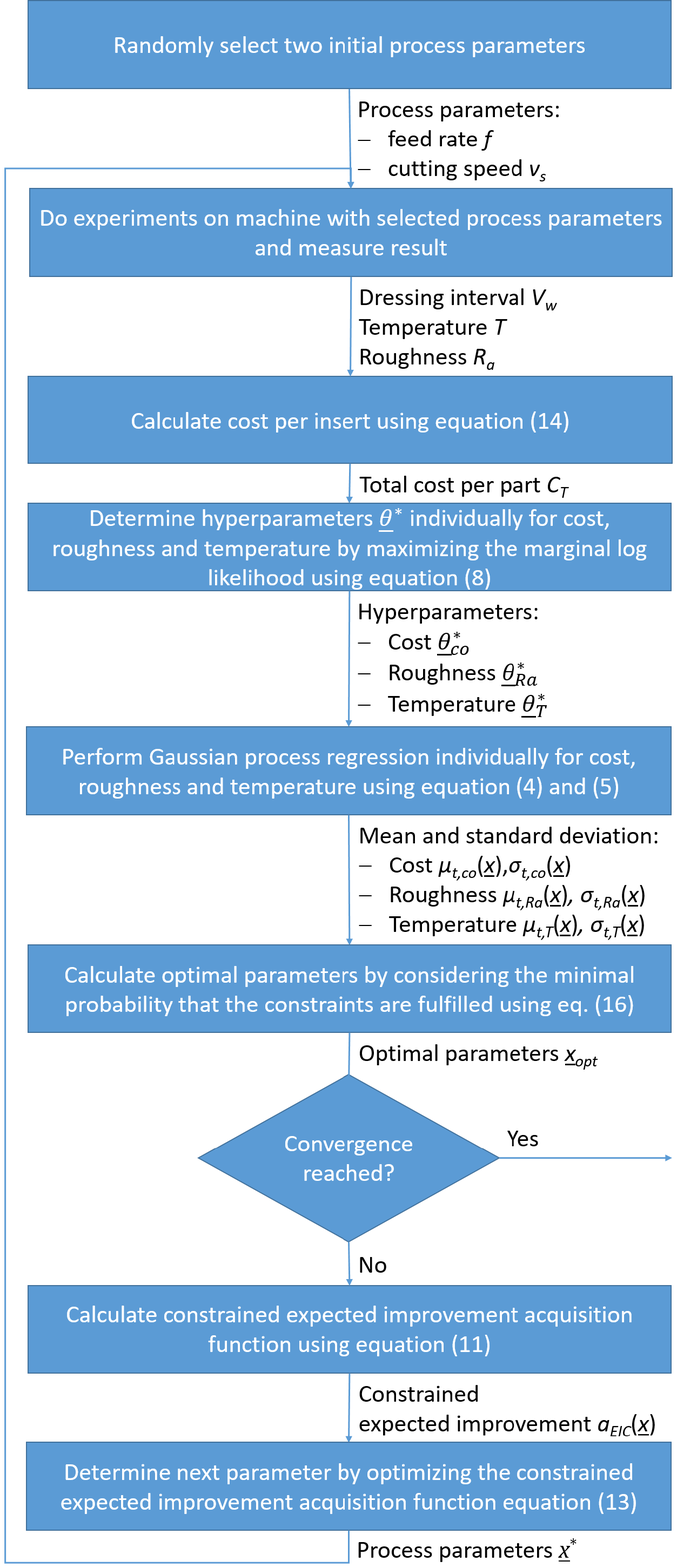}
\caption{Flow diagram of optimization implementation}
\label{fig:6}       
\end{figure}

The objective in this study is to find the process parameters minimizing the total production cost $C_T$ and to fulfill the maximum temperature constraint $T_{max}$ and the maximum surface roughness constraint $R_{a,max}$. 

\begin{equation}
\label{eqn:15}
\mathbf{x}_{min}=\mathrm{argmin}\,C_T(v_s,f)\,\,\mathrm{s.t.}\,\,\begin{array}{ll}
T(v_s,f)&<T_{max} \\
R_a(v_s,f)&<R_{a,max} \\
\end{array}
\end{equation}
In general, the optimization objective in machining is to minimize production costs, as applied in this study. However, other objectives such as energy consumption or environmental impact may also be considered. The algorithm was implemented in MATLAB using the GPML library \cite{Ref24} for Gaussian process regression. A flow diagram of the implementation is shown in Figure \ref{fig:6}. The optimization is started with two experiments at random process parameters within the input optimization domain. The optimization domain for the Bayesian optimization was set based on experience to a minimal feed rate of 10 mm/min and a maximum feed rate of 40 mm/min. The minimal cutting speed was set to 12 m/s and the maximum cutting speed was set to 30 m/s. In this way, a typical cutting speed range was covered.  After calculating the cost per part at the measured process parameter values and determining the hyperparameters, three Gaussian process regressions are calculated, to model the cost, the temperature, and the roughness.

Based on the obtained probabilistic predictions at different (not previously probed) values of the process parameters, the optimal parameters that fulfill (15) can be calculated based on the obtained probability distributions for the cost, temperature and roughness. Using stochastic models allows for probabilistic statements about the fulfillment of the constraint limits. As discussed in detail in \cite{Ref25}, the optimal parameters $\mathbf{x}_{opt} = (v_{s,opt}, f_{opt})$ are calculated as the parameters, which minimize the expected cost $\mu_{t,co}(v_s,f)$ after $t$ experiments calculated using equation (\ref{eqn:4}) and fulfill the roughness and temperature constraints with probabilities higher than user defined values $p_{Ra,min}$ and $p_{T,min}$. 
\begin{equation}
\label{eqn:16}
\mathbf{x}_{opt}=\mathrm{argmin}\,\mu_{t,co}(v_s,f)\,\,\mathrm{s.t.}\,\,\begin{array}{ll}
p_{f,T}(v_s,f)&\geq p_{T,min} \\
p_{f,Ra}(v_s,f)&\geq p_{Ra,min} \\
\end{array}
\end{equation}
The minimal probability that a constraint is fulfilled can be chosen freely and depends on the requirements. For example, many parts for the aviation industry will require a high minimal probability that the constraints are fulfilled. On the other hand if one produces disposable products it might be reasonable to accept a lower minimal probability that the constraints are fulfilled. The probability $p_{f,i}$  that the constraint $i$ is below its maximum allowed value $\lambda_i$ can be calculated for temperature $p_{f,T}$ and roughness $p_{f,Ra}$ by using equation (\ref{eqn:12}) and the maximum allowed temperature value $T_{max}$ = 585$^{\circ}$C and the maximum allowed roughness value $R_{a,max}$=230 nm (specified in section \ref{sec:32}). In general, it is possible to specify separate minimal probabilities for temperature and roughness. In this study, the parameters $p_{T,min}$ and $p_{Ra,min}$ were set equal for simplicity.

Afterwards the optimal parameters can be used to assess convergence as proposed in \cite{Ref18},
\begin{equation}
\label{eqn:17}
2\sigma_{t,co}(\mathbf{x}_{opt})<\epsilon
\end{equation}
where $\sigma_{t,co}$ is the predicted standard deviation of the cost at predicted optimal parameters $\mathbf{x}_{opt}$ after $t$ iterations and $\epsilon$ is the convergence limit. In this way, convergence is reached when 95\% of the predicted cost values at optimal parameters are within $\mu_c(\mathbf{x}_{opt})\pm \epsilon$. The convergence limit $\epsilon$ must be higher than the combined cost measurement and process uncertainty, otherwise convergence can never be reached. To increase the convergence robustness, full convergence is reached, when the stopping criterion is below 0.04 U over three consecutive iterations, the optimal feed rate change within an interval of 0.4~mm/min, and the optimal cutting speed change within an interval of 0.2 m/s. If convergence is not reached the next experimental parameters will be determined based on maximizing the constrained expected improvement acquisition function and the optimization continues.

\section{Results}
\label{sec:4}
\begin{figure}
\center
  \includegraphics[width=0.4\textwidth]{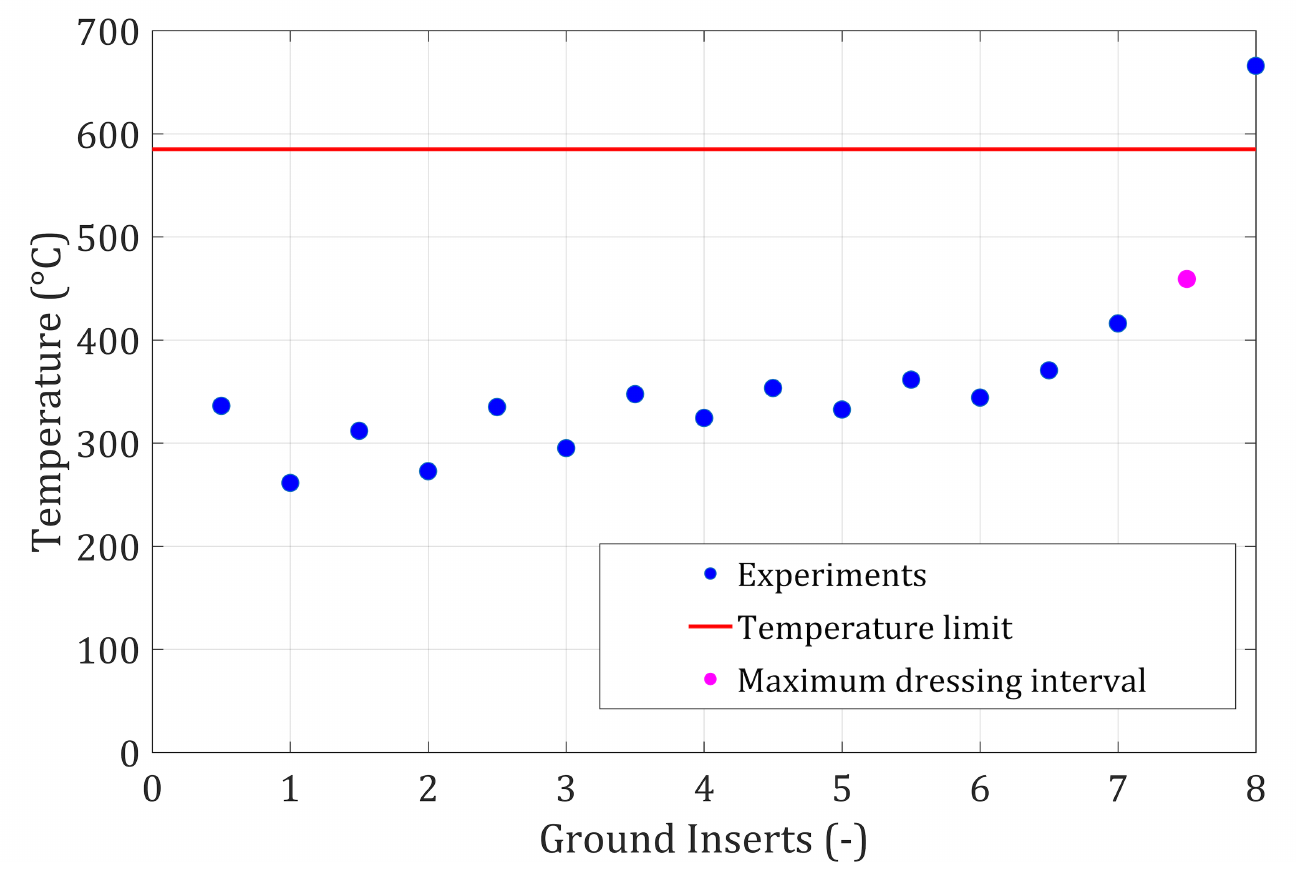}
\caption{Typical temperature measurement as a function of ground inserts for a cutting speed of 24.3 m/s and a feed rate of 11.7 mm/min. The last side which fulfills the temperature constraint is the first side of plate 8. Therefore, the maximum dressing interval is 7.5 inserts.}
\label{fig:7}       
\end{figure}

Figure \ref{fig:7} shows a typical temperature reading as a function of ground inserts for a full experimental run with constant input parameters. The grinding wheel was dressed before each experimental run, making it sharp and clean. In general, the temperature is strongly influenced by the accumulated removed material volume, where towards the end of an experimental run the temperature increases. The second side of the 8\textsuperscript{th} insert does not fulfill the constraint, resulting in grinding burn. Therefore, the measured maximum dressing interval for this parameter set is calculated as 7.5 inserts, which corresponds to an accumulated removed material volume $V_w$ of 2329.5 mm$^3$. Depending on the grinding task, considering workpiece handling time may lead to positive natural numbers as optimal dressing intervals instead of positive real numbers due to timesaving of dressing and workpiece handling parallelization. In this study, handling time of workpiece was neglected because it is considered very short compared to the dressing time and strongly depends on the used handling system and corresponding parameters.

\begin{figure}
\center
  \includegraphics[width=0.4\textwidth]{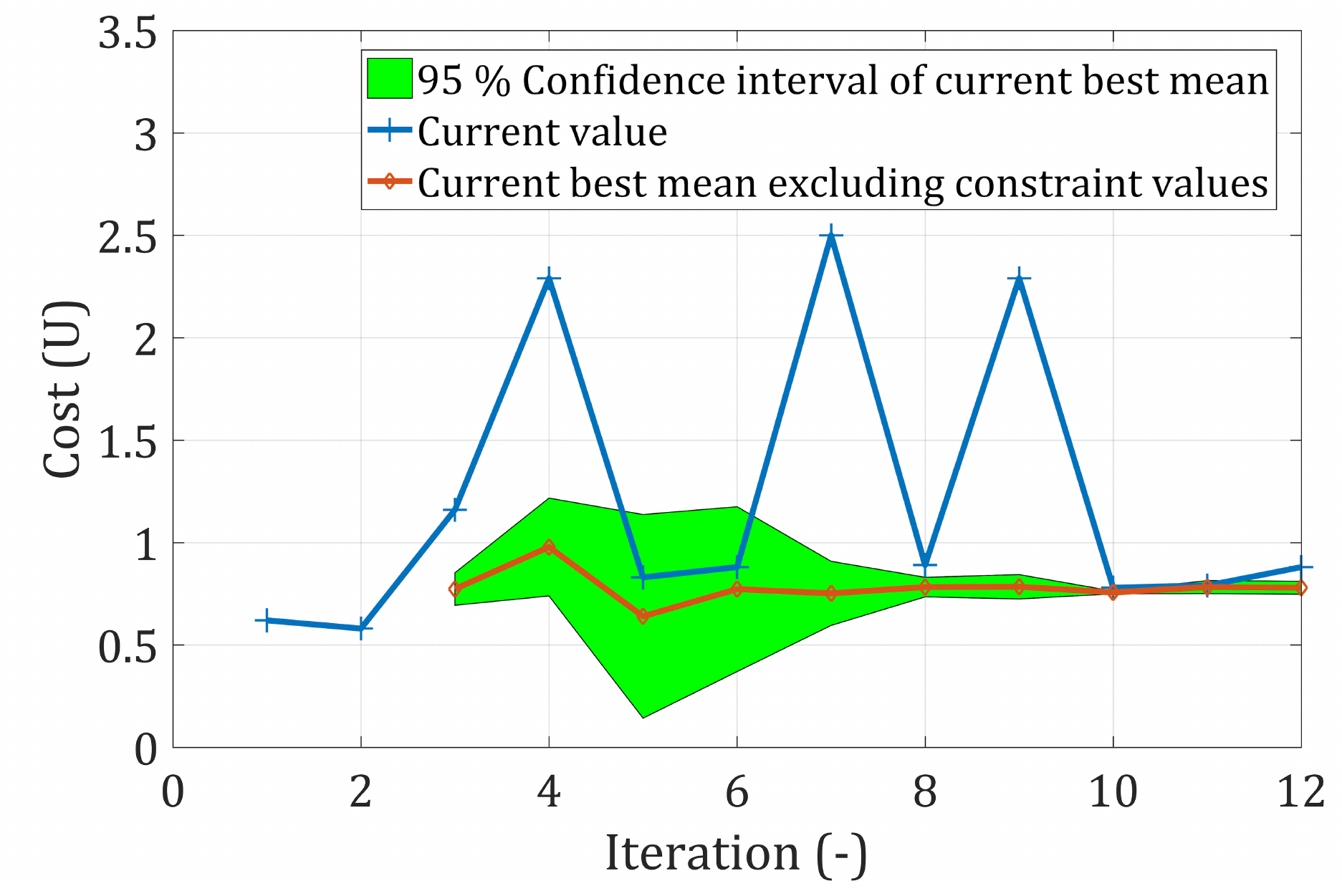}
\caption{Convergence of Bayesian optimization with a maximum allowed probability of quality defects $p_{Ra,min}$=$p_{T,min}$=0.5}
\label{fig:8}       
\end{figure}

\begin{figure}
\center
  \includegraphics[width=0.4\textwidth]{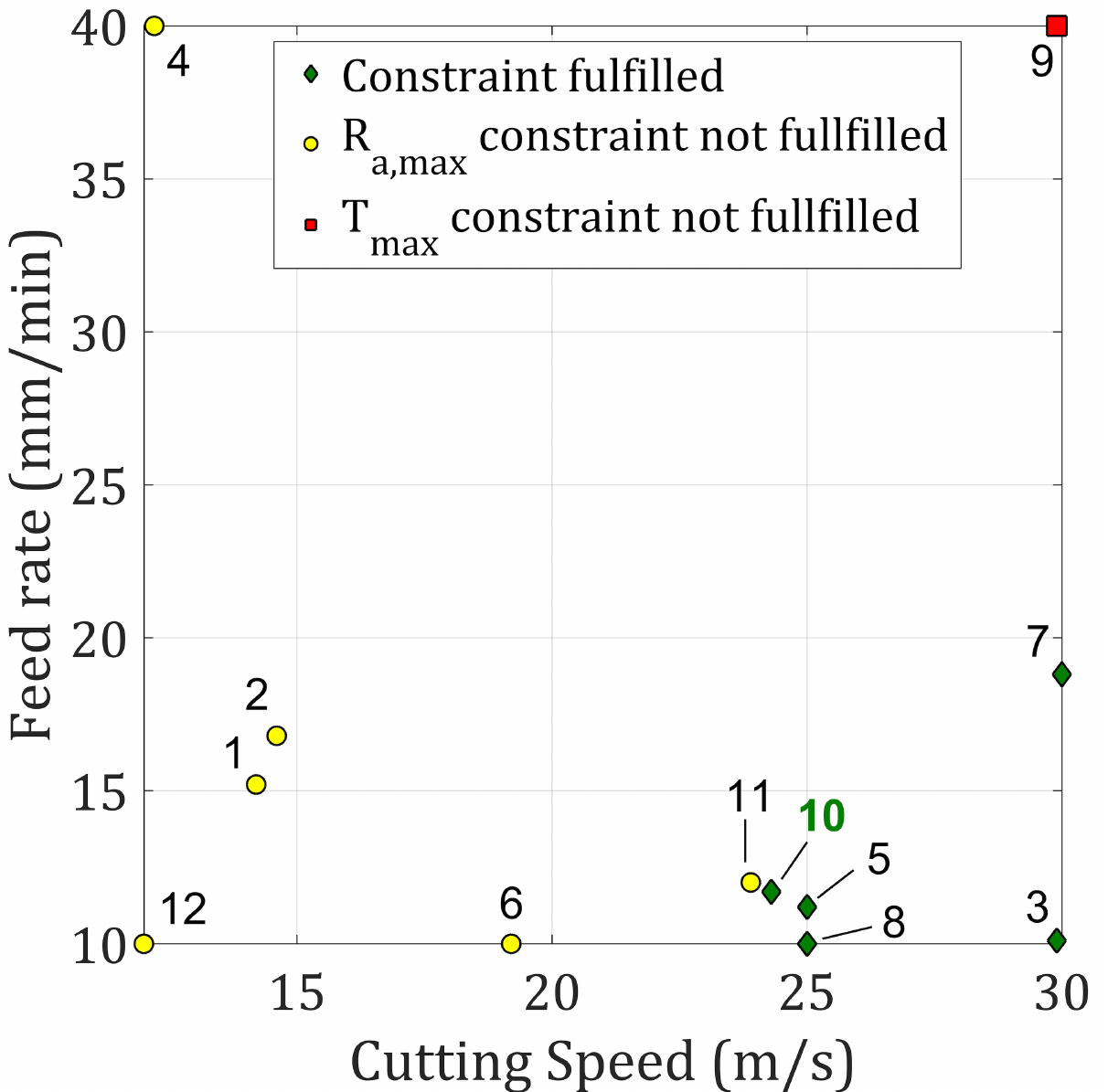}
\caption{Conducted experiments after 12 iterations. Experiment with lowest cost and fulfilled constraints was 10\textsuperscript{th} experiment}
\label{fig:9}       
\end{figure}

Figure \ref{fig:8} and Figure \ref{fig:9} present the on-machine optimization. Convergence was judged for a minimal allowed probability of quality defects of 0.5 ($p_{Ra,min}=p_{T,min}=0.5$). The first two experiments are conducted at random points in the parameter space to initialize the optimization. The first and the second experiment do not fulfill the constraints, therefore the algorithm does not approach convergence after these experiments. However, the proposed trial for the third measurement is at high cutting speeds and low feed rates, which reduces both surface roughness and temperature. The third measurement fulfills the constraints. Until the 8\textsuperscript{th} iteration the uncertainty of the prediction is reduced drastically. After the 8th iteration, the uncertainty remained low until full convergence is reached after 12 iterations. Experimentally, the best parameters are obtained at the 10\textsuperscript{th} iteration with a cutting speed of 24.3 m/s, a feed rate of 11.7 mm/min, and a dressing interval of 7.5 inserts. After the 10\textsuperscript{th} iteration, a small increase in the cost and the uncertainty is observed, however within the convergence limits. The blue line on Figure \ref{fig:8} shows the current measured cost values. Experimental points explored due to high model uncertainty often result in high costs, for example iterations 4, 7, and 9. In such cases, often the constraints are not fulfilled.

\begin{figure}
\center
  \includegraphics[width=0.6\textwidth]{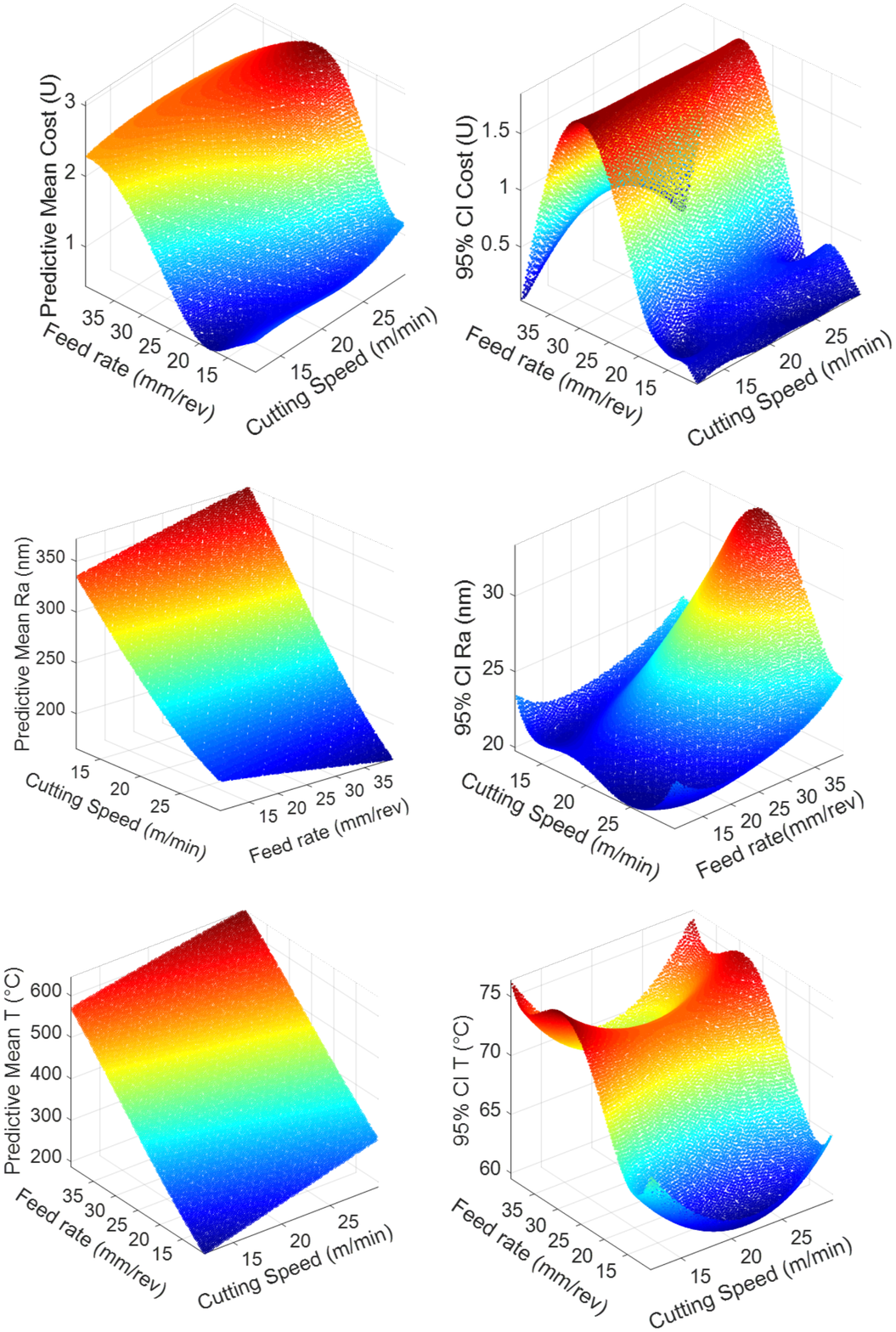}
\caption{Predicted mean cost (top left), cost confidence interval (top right), predicted mean roughness (middle left), roughness confidence interval (middle right), predicted mean temperature (bottom left), and temperature confidence interval (bottom right) after 12 iterations.}
\label{fig:10}       
\end{figure}

Figure \ref{fig:10} shows the predicted mean and confidence interval after 12 iterations for cost, roughness and temperature. The predicted cost is high for low feed rates because the operation is time consuming. On the other hand, higher feed rates lead to shorter dressing intervals, which makes redressing expensive. Optimal parameters are a trade-off between these two costs. After 12 experiments, the model uncertainty in the high feed rate range (above 25~m/min) is higher than the uncertainty corresponding to low feed rate. Considering the high predicted costs for these parameters, the uncertainty is still moderate. The roughness mostly depends on the cutting speed. An increase in cutting speed reduces the surface roughness whereas the feed rate only influences the roughness minimal. This is in line with the findings of the preliminary study \cite{Ref21} for cup wheel grinding with the same setup, where  an extensive series of experiments was conducted. The predicted uncertainty of the roughness is between 19.7 and 33.4 nm, which is 9 and 15\% of the maximum allowed roughness. An increase in feed rate or cutting speed leads to higher temperatures, again confirming the results reported in \cite{Ref21}. The predicted temperature uncertainty is between 59.5 and 76.5$\circ$C which is between 10 and 13\% of the maximum allowed temperature. The temperature uncertainty is higher for high feed rates because only limited number of experiments were conducted at high feed rates due to high total costs at high feed rates. In such cases, the recommended process parameter configuration by the BO algorithm for the next trials moves to process parameters corresponding to lower costs, and the more unfavorable regions of the parameter space are not extensively explored. 

Table \ref{tab:2} shows optimal parameters for several different values of $p_{Ra,min}$ and $p_{T,min}$, computed from equation \ref{eqn:16} after 12 iterations. The algorithm recommends an increase in cutting speed when the minimal probabilities that the constraints are fulfilled $p_{Ra,min}$ and $p_{T,min}$ are increased, without a large modification of the feed. An increase in cutting speed leads to lower surface roughness and less probability to violate the surface roughness constraint. As expected, increasing the minimal probability that the constraints are fulfilled results in higher production costs. 
\begin{table}
\centering
\caption{Optimal predicted parameters after 12 iterations with different minimal probabilities that the temperature and roughness constraints are fulfilled.  The predicted optimal costs and the predicted 95\% confidence interval of the optimal costs is also given.}
\label{tab:2}       
\begin{tabular}{lll}
\hline\noalign{\smallskip}
Minimal probability that\\ constraints are fulfilled (\%) & Optimal  parameters & Costs (U)  \\
\noalign{\smallskip}\hline\noalign{\smallskip}
50   & $f$ = 12.0 mm/min, $v$ = 23.8 m/s & 0.78$\pm$0.04 \\
84.1 & $f$ = 11.8 mm/min, $v$ = 25.2 m/s & 0.83$\pm$0.05 \\
97.7 & $f$ = 11.4 mm/min, $v$ = 26.8 m/s & 0.92$\pm$0.07 \\
99.9 & $f$ = 11.6 mm/min, $v$ = 28.8 m/s & 1.08$\pm$0.13 \\
\noalign{\smallskip}\hline
\end{tabular}
\end{table}
\section{Conclusion}
\label{sec:5}
In this study, Bayesian optimization with on-machine measurements was used to obtain optimal parameters for grinding within a few iterations, while fulfilling quality and safety constraints. Gaussian process regression models were successfully applied to model costs, roughness, and temperature of the grinding process. The proposed approach is suitable to consider the probability that the constraints are fulfilled in the calculation of optimal parameters. In future, additional input parameters such as dressing feed rate, dressing wheel speed, and grinding wheel oscillation frequency might be included in the optimization. Furthermore, the optimization will benefit from additional sensor feedback such as grinding wheel wear sensors and integrated cutting edge roughness measurements.

%
\section*{Conflict of interest}
Agathon AG filed a patent on parts of this work. 
%


%
%


\bibliography{bib1.bib}
\bibliographystyle{IEEEtran}
\end{document}